\documentstyle[prl,aps,epsf]{revtex}
\newcommand{\gve}{\gamma_\varepsilon}
\newcommand{\gv}{\gamma_v}
\newcommand{\ome}{\omega_E}
\newcommand{\os}{\omega_s}
\newcommand{\Tsym}{\frac{2n+1}{2}T}
\newcommand{\Vavg}{\langle V \rangle}
\newcommand{\Eavg}{\langle E_{sc} \rangle}
\newcommand{\Esavg}{\langle eaE_{sc}/\hbar\ome \rangle}
\newcommand{\Uavg}{\langle U \rangle}
\begin{document}
\draft
\title{Symmetry-Breaking and Chaos in Electron Transport in Semiconductor 
       Superlattices}
\author{Kirill N. Alekseev$^{1,2,3}$, Ethan H. Cannon$^{1}$, Jonathan C. 
McKinney$^{1}$, Feodor V. Kusmartsev$^{4}$\cite{landau}, and D.K. Campbell$^1$ 
\\
$^1$Department of Physics, University of Illinois at 
Urbana-Champaign, 1110 West Green St., Urbana, IL 61801, USA\\
$^2$Department of Physical Sciences, Box 3000, University of Oulu FIN-90014, 
Finland\\
$^3$Theory of Nonlinear Processes Laboratory, Kirensky Institute of
Physics, Krasnoyarsk 660036, Russia\\
$^4$Department of Physics, Loughborough University, Loughborough LE11 3TU, UK}
\maketitle
\begin{abstract}
We study the motion of ballistic electrons in a single miniband of a 
semiconductor superlattice driven by a terahertz laser polarized along 
the growth direction. We work in the semiclassical balance-equation model, 
including different elastic and inelastic scattering rates, and incorporating 
the self-consistent electric field generated by electron motion. We explore 
regions of complex dynamics, which can include chaotic behavior and 
symmetry-breaking. Finally, we estimate the magnitude of the DC current and 
voltage that spontaneously appear in regions of broken-symmetry for parameters 
characteristic of real semiconductor superlattices.
\end{abstract}

\section{Introduction}

Since the seminal work of Esaki and Tsu [1] predicting negative 
differential conductivity in the current-voltage characteristics of
a semiconductor superlattice (SSL), SSLs have been the subject of intense
theoretical and experimental study.  SSLs consist of thin layers of alternating
semiconductors -- typically 10-100 layers ranging from 10 to 500 \AA\ in 
thickness -- whose different band structures create a periodic potential 
along the growth axis, essentially forming a one-dimensional, artificial 
crystal. The superlattice period can be controlled during the growth process and 
is generally much larger than the period of the constituent materials. 
Consequently the conduction and valence bands break into narrow allowed and 
forbidden regions termed minibands, with typical widths ranging from 1 to 100 
meV. For an introduction to SSL design and properties, see reference [2].
\par
Recent experiments on SSLs in the presence of high-frequency fields have 
revealed many novel, nonlinear effects, including photon-assisted tunneling [3],
a reduction of the current during laser irradiation [4],
absolute negative conductivity [5], resonant increases in the current [6],
and, most recently, deterministic chaos in the sequential resonant tunneling 
regime [7]. Our previous work [8] suggests that chaos should also occur in the 
miniband regime for a SSL driven by an AC field.
In this brief note, we recall some of the principal results of
that work and report our current studies of symmetry-breaking
effects in electron transport. Using numerical simulations we
establish the possibility of spontaneous DC current and voltage
generation in response to a purely AC drive. Our results cover the
whole range of frequency/amplitude of the applied field, extending
a recent investigation of the high frequency, strong field limit [9].

\section{Balance Equations}

We consider electrons subject to an external field 
$E_{ext}=E_0\cos(\Omega t)$ using the tight-binding
dispersion relation, $\epsilon(k)=\frac{\Delta}{2}[1-\cos(ka)]$,
for SSL period $a$, minibandwidth $\Delta$, electron energy $\epsilon$
and crystal momentum $k$. We assume that the frequency of the external
field is not large enough to support Zener tunneling to higher minibands,
and that the electric field is uniform across the SSL.
To model the electron dynamics, we generalize the semiclassical balance
equations of Ignatov et al. [10] by including the self-consistent electric 
field resulting from electron motion, obtaining [8],
\begin{eqnarray}
\dot{V}& = & -eE_{tot}(t)/m(\varepsilon)-\gv V \label{eq:Vt} \\
\dot{\varepsilon}&=& -eE_{tot}(t)V-\gve(\varepsilon-\varepsilon_0) \\
\dot{E}_{sc}&=& -4\pi j/\kappa-\alpha E_{sc} \label{eq:Et}
\end{eqnarray}
where $E_{tot}=E_{ext}+E_{sc}$ is the total electric field, including the 
self-consistent contribution; $j=-eNV$ is the current density;
$V(t)$ and $\varepsilon(t)$ are the average electron velocity and energy,
respectively;
$\gv$, $\gve$, and $\alpha$ are phenomenological relaxation rates for the
average velocity, average energy and self-consistent field, respectively;
$\kappa$ is the average dielectric constant for the SSL;
$\varepsilon_0$ is the temperature-dependent equilibrium energy;
$N$ is the carrier concentration, assumed uniform across the SSL; and
$m(\varepsilon)=\frac{m_0}{1-2\varepsilon/\Delta}, 
m_0=\frac{2\hbar}{\Delta a^2}$, is the effective mass of the electrons.
\par
The balance equations remain invariant under the symmetry transformation:
\begin{equation}
t \rightarrow t+\Tsym, \ V\rightarrow -V, \ E_{sc}\rightarrow -E_{sc}, 
\label{eq:symm}
\end{equation}
where $T$ is the period of the applied field and $n=0,1,2, \cdots$; 
in other words, if $\{V(t), \varepsilon(t), E_{sc}(t)\}$ 
satisfy (\ref{eq:Vt})--(\ref{eq:Et}), 
then so do $\{-V(t+\Tsym), \varepsilon(t+\Tsym), -E_{sc}(t+\Tsym)\}$.  
Physically, this represents the lack of preferred spatial direction 
for a SSL without DC bias; under this transformation
the electron velocity and the self-consistent field
reverse direction, but the electron energy remains unaltered.  
\par
Symmetric {\em solutions} also remain invariant under transformation 
(\ref{eq:symm}); hence
$V(t+\Tsym)=-V(t)$ and $E_{sc}(t+\Tsym)=-E_{sc}(t)$.  For such cases the 
time-averages of both the velocity and the self-consistent field are zero.  
However there may also exist solutions which do not satisfy this symmetry, 
for which $V$ and $E_{sc}$ have non-zero average values; 
these symmetry-broken solutions correspond to the spontaneous
generation of a DC current and bias.  
\par
In the limiting case $\gv=\gve=0$, (\ref{eq:Vt})--(\ref{eq:Et}) 
can be transformed to the equation describing the Resistively Shunted 
Junction (RSJ) model of a Josephson junction (JJ) [8,11]
This model possesses a thoroughly studied symmetry analagous to (\ref{eq:symm}) 
[12,11]. When biased by an AC current of frequency $\Omega$,
the JJ may develop a DC component to its voltage difference satisfying
the phase-locking condition $\Uavg \propto n\Omega$, where $n$ is an integer, 
$U$ is the voltage across the JJ, and angular brackets indicate time-averages 
[12].

\section{Numerical Studies}

In our earlier work [8], we showed that the balance equations give rise to
chaotic behavior for a wide range of damping and driving parameters.
Readers should consult this reference for a full description of the
numerics and scalings used in the simulations.  Briefly, however, there
are three important frequencies: the drive frequency, $\Omega$; the Stark
frequency of the driving field, $\os=eaE_0/\hbar$; and the 
frequency of the cooperative oscillations associated with the self-consistent
field, $\ome=\sqrt{2\pi e^2 N a^2 \Delta/\kappa\hbar^2}$ which is usually in
the terahertz range.  As it depends only on
SSL material parameters, we use $\ome$ as the scale of inverse time, and
throughout the rest of this article we consider 
the dimensionless frequencies $\Omega/\ome$, $\os/\ome$, $\gv/\ome$,
$\gve/\ome$ and $\alpha/\ome$ without explicitly showing the scaling.
\par
As in many studies of the JJ [12,11], we examine the dynamics
for fixed damping while varying the parameters of the applied field.  
Figures 1a and 1b indicate the region of chaos and symmetry-breaking, 
respectively.
Note the similar general shape of these regions; further work shows that
for greater damping the region of chaos diminishes while the region of
symmetry-breaking retains this general shape [8,13].
\par
Fig. 1c plots the time-averaged self-consistent field, $\Esavg$,
as a function of external frequency, $\os$, for the frequency $\Omega=0.9$.  
Phase-locked symmetry-breaking where $\Esavg \approx n\Omega/\ome$ for 
$n=\pm1,\pm2$ is clearly visible.  A careful examination reveals that $\Esavg$
also depends weakly on $\os$; therefore the phase-locking relation is
only approximately satisfied.
\par
From (\ref{eq:Et}) it immediately follows that $\Vavg\propto\Eavg$ for
non-chaotic solutions.  Consequently 
symmetry-breaking should be experimentally observable as a DC current and bias
across an unbiased SSL illuminated by a terahertz laser.  For representative
SSL parameters of $a=100 \AA$, $N=3\times 10^{15} cm^{-3}$, $\Delta=22 meV$,
and $\epsilon=13$ [6],
$\ome=3.3\times 10^{12} s^{-1}$.  Taking $\Esavg_{max}=1.5$ yields
a DC bias per SSL period of $3mV$.  
Furthermore, a lateral area of $8 \mu m^2$ would result 
in a current of $1\mu A$.  Both these magnitudes should be experimentally
detectable.

\section{Conclusion}

We have studied the complex dynamics of semiclassical balance equations that 
model the motion of electrons in the miniband of a semiconductor superlattice
exposed to an external ac field.  The equations include the self-consistent
field created by electron motion.  Numerical simulations indicate the 
possibility of broken-symmetry solutions, which occur in roughly the same
laser frequency and electric field ranges as chaotic dynamics.  The 
symmetry-breaking engenders a DC bias and current that could be experimentally 
detectable.
\par
We would like to contrast this work with recent theoretical [14]
and experimental [7] 
work predicting and observing chaos for SSLs in the 
incoherent tunneling regime when the electric field breaks into domains 
of different magnitudes [2].
Our model considers the coherent motion of electrons within a miniband
in a spatially homogenous electric field.  A complete theoretical 
understanding of when the applied field breaks into domains has not yet been
attained, so it is encouraging that chaos is predicted in quite different
models applicable in different regimes of SSL design and applied field
characteristics.
\par
We are especially grateful to Gennady Berman for his advice and continuing
collaboration on many aspects of chaos in mesoscopic systems. We
thank Sergei Turovets for a discussion on the notation of 
symmetry-breaking in dynamical systems.
K.N.A. thanks the Department of Physics at The University of Illinois at
Urbana-Champaign for hospitality. This work was partially supported by
Linkage Grant No. 93-1602 from the NATO Special Programme Panel on 
Nanotechnology. K.N.A. was supported by INTAS(94-2058) and KRSF(6F0030).
E.H.C. thanks the U.S. Department of Education for support by a GAANN 
Fellowship (DE-P200A40532), and J.C.M. thanks the U.S.-NSF for support 
under its REU program (NSF Grant No. PHYS93-22320).

\begin{figure}
\epsfxsize=8cm
\hspace{3cm} 
\epsfbox{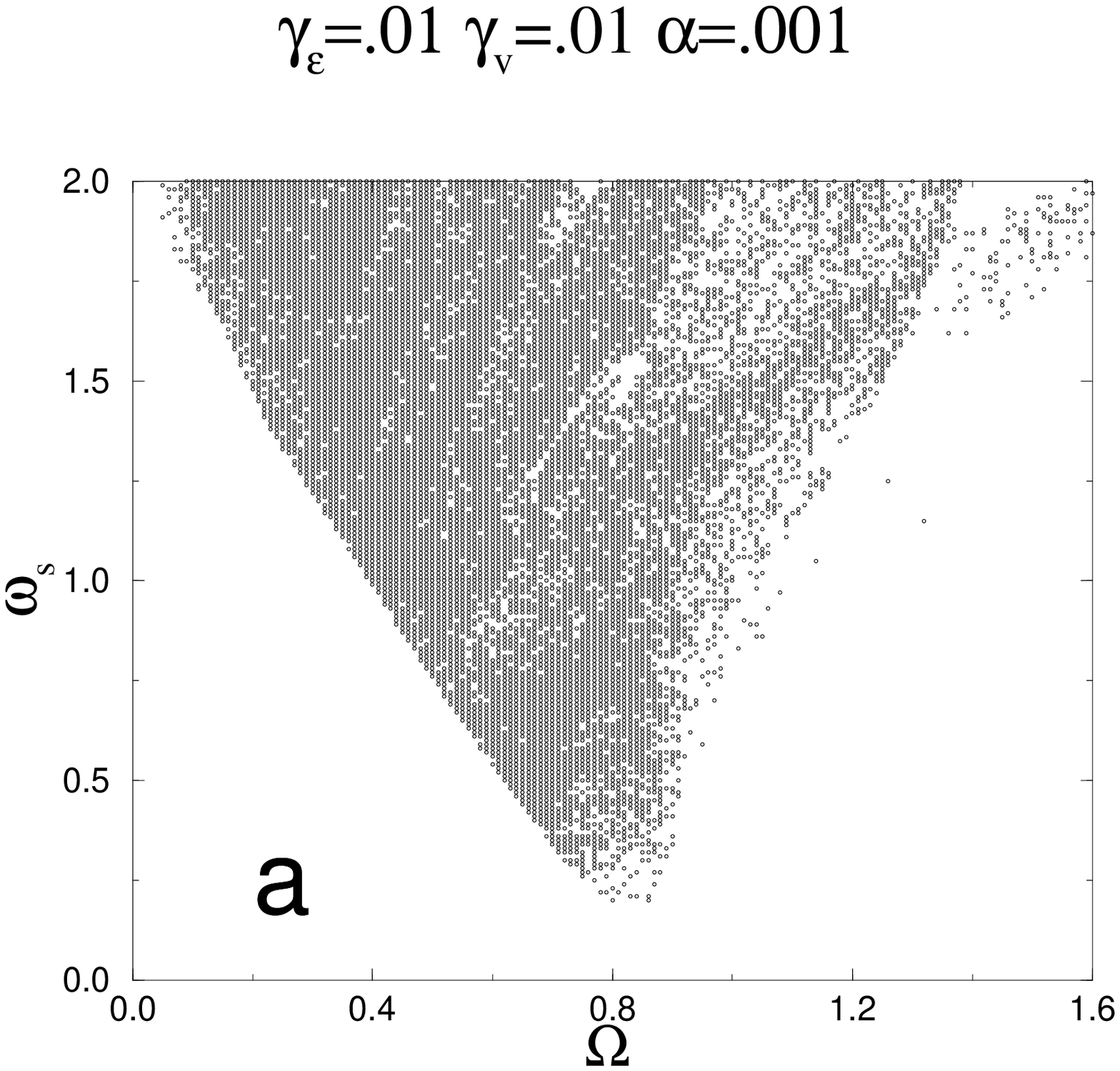}
\end{figure}
\begin{figure}
\vspace{0.5cm}
\epsfxsize=8cm
\hspace{3cm} 
\epsfbox{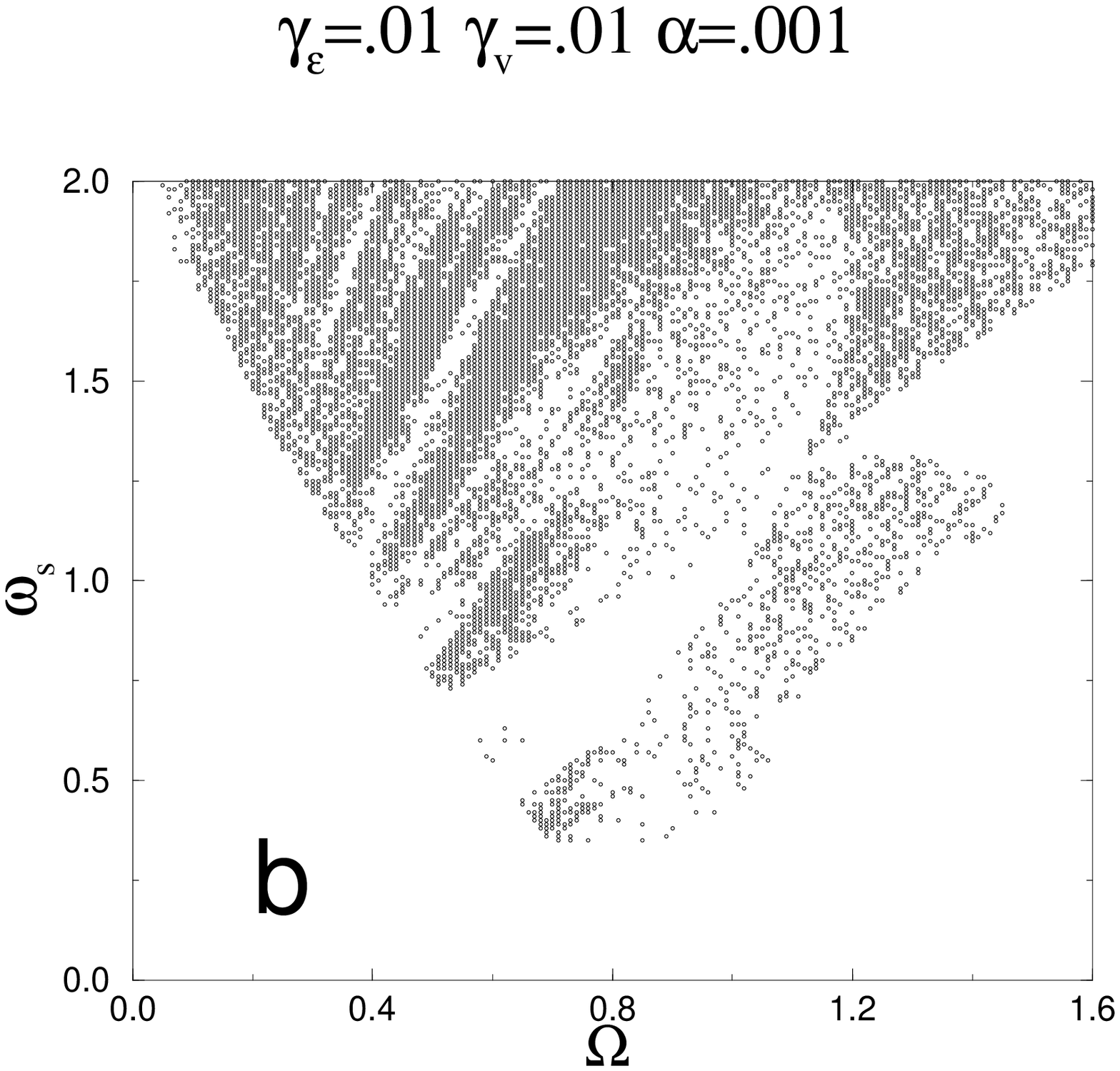}
\end{figure}
\begin{figure}
\vspace{0.5cm}
\epsfxsize=8cm
\hspace{3cm} 
\epsfbox{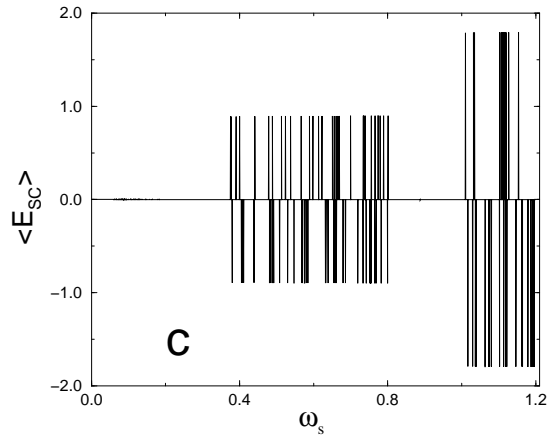}
\caption{For a weakly damped SSL, chaotic (dark symbols) and regular (white 
areas) regions (a); symmetry-broken (dark symbols) and symmetric (white areas) 
regions (b); time-averaged self-consistent field as a function of applied field 
strength (c)}
\end{figure}

\end{document}